\documentclass[a4paper, 12pt, oneside]{article}
\usepackage[cp1251]{inputenc}
\usepackage[english, russian]{babel}
\usepackage{amsmath, latexsym,  amssymb, array, graphics, amsfonts, amsthm, bm}


\usepackage{indentfirst}
\makeatother

\begin{document}
\renewcommand{\abstractname}{\ }
\renewcommand{\refname}{\begin{center} REFERENCES\end{center}}
\thispagestyle{empty}
\large

\begin{center}
\bf Maxwell problem about thermal sliding of rarefied gas along plate plane
\end{center}\medskip

\begin{center}
  \bf  A. V. Latyshev\footnote{$avlatyshev@mail.ru$},
  A. A. Yushkanov\footnote{$yushkanov@inbox.ru$} and
  E. E. Korneeva\footnote{$ee-korneeva@yandex.ru$}
\end{center}\medskip

\begin{center}
{\it Faculty of Physics and Mathematics,\\ Moscow State Regional
University,  105005,\\ Moscow, Radio str., 10--A}
\end{center}\medskip

MSC2010: 82C40, 82B40, 80A20, 80A99

\begin{abstract}
One of classical boundary problems of the kinetic theory
(a problem about thermal sliding) of the rarefied gas along a flat firm
surface is considered. Kinetic Boltzmann equation with model
integral of collisions BGK (Bhatnagar, Gross, Krook) is used. As boundary
conditions the boundary Maxwell conditions (mirror-diffuse) are used.
The generalized method of a source is applied
to the problem decision. Comparison with earlier received results is spent.
\end{abstract}

\medskip

{\bf Key words:} thermal sliding,  rarefied gas, accommodation coefficient,  kinetic equation,
Fredholm equation, distribution function, Neumann series.
\medskip

PACS numbers: 51. Physics of gases, 51.10.+y  Kinetic and
transport theory of gases.

\begin{center}
{\bf Introduction}
\end{center}

Maxwell was the first who has paid attention to the movement of the rarefied gas
under the influence of heterogeneous temperature distribution \cite{1}-\cite{3}. A problem
about the thermal sliding of gas along a surface (not necessarily flat)
causes constant interest (see, for example, \cite{4}-\cite{12}. It  is related  both to
rarely  theoretical interest and with numerous applications in area   of
aerodynamics and physics of the aero dispersible systems.
The most complete review of works in this direction is presented in work \cite{4}
and the monograph \cite{5}. We note a number of works \cite{6}-\cite{9}, mirror-boundary
conditions were examined in that.  A great contribution to the study of thermal
sliding brought  Loyalka S. K. \cite{7}-\cite{9}.
In our works \cite{10}-\cite{15} were worked out approximation
\cite{10}-\cite{12} and exact \cite{13,14} methods of solution of
boundary value problems for model kinetic equations.
In works \cite{10}, \cite{11} the coefficient of thermal sliding is app\-ro\-xi\-ma\-ted by
fractional rational functions. Analytical solution of the problem on thermal
sliding for a gas with  frequency of collisions pro\-por\-tional to the modulus
of the speed of molecules, and diffuse boundary value problems, it was got
in  \cite{15}. Then in works \cite{16}-\cite{18} thermal sliding was
considered for quantum gases.

\begin{center}
{\bf 1. Statement problem}
\end{center}

Let the rarefied gas fills a half-space $x>0$   and moves along an axis $y$.
Far from a surface ($y, z$)  the logarithmic gradient of temperature
$$
g_T=\left( \dfrac{d \ln T(y)}{d y}\right)_{x= +\infty}
$$
is set.

It is required to find the speed of thermal sliding  $u_{sl}=u_y(+\infty)$ and
distribution of mass speed of gas $u_y=u_y(x)$  in a half-space.

In work \cite{12} it is shown that if to search the  function of distribution in the form
$$
f(t,{\bf r},{\bf v})=f_M(v)(1+h(t,{\bf r},{\bf v})),
$$
where
$$
f_{M}(\mathbf{r},\mathbf{v},t)=n(y)\Big(\dfrac{m}{2\pi kT(y)}\Big)^{3/2}\exp
\Big[-\dfrac{m}{2kT(y)}\mathbf{v}^2\Big],
$$
$$
T(y)=T_0(1+g_Ty),
$$
then function  $h(x,\mu)\; (\mu=C_x)$  satisfies to the nonhomogeneous  kinetic
equation
$$
\mu\dfrac{\partial h}{\partial x_1}+G_T\Big(\mu^2-\dfrac{1}{2}\Big)+h(x_1,\mu)=
\dfrac{1}{\sqrt{\pi}}\int\limits_{-\infty}^{\infty}e^{-{\mu'}^2}h(x_1,\mu')d\mu.
\eqno{(1.1)}
$$

Here and below
$$
G_T=\Big(\dfrac{d\ln T}{dy_1}\Big)_{x_1=+\infty},\qquad x_1=\nu \sqrt{\beta}x, \qquad
y_1=\nu \sqrt{\beta}y,
$$
$$
\beta=\dfrac{m}{2kT_0}, \qquad U_{sl}=\sqrt{\beta}u_{sl},
$$
$\nu$ is  the frequency of collisions, $x_1, y_1$ is the dimensionless coordinates,
$k$ is the Boltzmann constant.

Further we will consider that $x_1 \equiv x, y_1\equiv y$.

We will consider that molecules are reflected from a wall mirror-diffuse
$$
f(t,+0, \mathbf{v})=qf_0(v)+(1-q)f(t,+0,-v_x,v_y, v_z), \quad v_x>0,
\eqno{(1.2)}
$$
where $q$   is the accomodation coefficient (coefficient of diffusively),
$0\leqslant q \leqslant 1$.

At the $q=1$ condition (1.2) is a condition of diffuse reflection, at $q=0$
is the condition of a specular reflection.

For function $h(x,\mu)$  the condition (1.2) passes into the  condition
$$
h(0,\mu)=(1-q)h(0,-\mu),\qquad \mu>0.
\eqno{(1.3)}
$$

Far from a wall function of distribution turns into the Champen---Enskog distribution
$$
h(\infty,\mu)=2U_{sl}-G_T\Big(\mu^2-\dfrac{1}{2}\Big).
\eqno{(1.4)}
$$

We will designate further
$$
h(x,\mu)+G_T\Big(\mu^2-\dfrac{1}{2}\Big)=\psi(x,\mu).
$$

Then the equation (1.1) passes into the homogeneous equation
$$
\mu\dfrac{\partial \psi}{\partial x}+\psi(x,\mu)=\dfrac{1}{\sqrt{\pi}}
\int\limits_{-\infty}^{\infty}e^{-t^2}\psi(x,t)\,dt,
\eqno{(1.5)}
$$

Boundary conditions (1.3) and (1.4) pass into the following
$$
\psi(0,\mu)=qG_T\Big(\mu^2-\dfrac{1}{2}\Big)+(1-q)\psi(0,-\mu),\qquad \mu>0,
\eqno{(1.6)}
$$
and
$$
\psi(0,\mu)=2U_{sl}.
\eqno{(1.7)}
$$

Further we will put
$$
\psi(x,\mu)+2U_{sl}+h_c(x,\mu).
$$

Thus function   satisfies to the equation (1.5)
$$
\mu\dfrac{\partial h_c}{\partial x}+h_c(x,\mu)=\dfrac{1}{\sqrt{\pi}}
\int\limits_{-\infty}^{\infty}e^{-t^2}h_c(x,t)\,dt,
\eqno{(1.8)}
$$
and to boundary conditions
$$
h_c(0,\mu)=-2qU_{sl}+qG_T\Big(\mu^2-\dfrac{1}{2}\Big)+(1-q)h_c(0,-\mu),\qquad \mu>0,
\eqno{(1.9)}
$$
and
$$
h_c(\infty,\mu)=0.
\eqno{(1.10)}
$$

\begin{center}
{\bf 2. Inclusion of boundary conditions in the kinetic equation}
\end{center}

We will continue function of distribution to the interfaced half-space $x<0$
by symmetric character
$$
h(x,\mu)=h(-x,-\mu), \qquad \mu>0.
$$

At such  extension the function $h_c(x,\mu)$   in negative half-space $x<0$
satisfies to equation (1.8)  again  and  to the boundary conditions
$$
h_c(-0,\mu)=$$$$=-2qU_{sl}+qG_T\Big(\mu^2-\dfrac{1}{2}\Big)+(1-q)h_c(-0,-\mu),\qquad
\mu<0,
\eqno{(2.1)}
$$
and
$$
h_c(-\infty,\mu)=0.
\eqno{(2.2)}
$$

We will include boundary conditions (1.9), (1.10) and (2.1) (2.2) in the kinetic
equation in the form of a source, which member contains Dirac's delta function\medskip
$$
\mu \dfrac{\partial h_c}{\partial x}+h_c(x,\mu)=$$$$=2U_c(x)+
|\mu|\Big[-2qU_{sl}+qG_T\Big(\mu^2-\dfrac{1}{2}\Big)-q h_c(\mp 0,-\mu)\Big]
\delta(x),
\eqno{(2.3)}
$$\medskip
where $\delta(x)$  is Dirac's delta function, and
$$
2U_c(x)=\dfrac{1}{\sqrt{\pi}}\int\limits_{-\infty}^{\infty}e^{-{\mu'}^2}h_c(x,\mu')d\mu'.
\eqno{(2.4)}
$$

We will notice that  $U_c(x)=0$  satisfies to the boundary conditions
$$
U_c(\pm \infty)=0.
\eqno{(2.5)}
$$

For the equation (2.3) at $x>0, \mu<0$,
we get the solution satisfying to the  boundary conditions (1.9) and (2.5)
$$
h_c^+(x,\mu)=-\dfrac{1}{\mu}\exp(-\dfrac{x}{\mu})
\int\limits_{x}^{+\infty} \exp(+\dfrac{t}{\mu})2U_c(t)\,dt.
\eqno{(2.6)}
$$

Similarly, at  $x<0,\,\mu>0$ we find
$$
h_c^-(x,\mu)=\dfrac{1}{\mu}\exp(-\dfrac{x}{\mu})
\int\limits_{-\infty}^{x} \exp(+\dfrac{t}{\mu})2U_c(t)\,dt.
\eqno{(2.7)}
$$

We will rewrite the equation (2.3), and we will replace in it the last member,
using continuation of function $h(x,\mu)$ on the interfaced half-space and
equalities (2.6) and (2.7). On this way we come to the following equation
$$
\mu\dfrac{\partial h_c}{\partial x}+h_c(x,\mu)=$$$$=2U_c(x)+
|\mu|\Big[-2qU_{sl}+qG_T\Big(\mu^2-\dfrac{1}{2}\Big)-qh_c^{\pm}(0,-\mu)\Big]\delta(x).
\eqno{(2.8)}
$$

Here
$$
h_{c}^{\pm}(0,\mu)=-\dfrac{1}{\mu}e^{-x/\mu} \int\limits_{0}^{\pm
\infty}e^{t/\mu}2U_c(t)dt.
$$

We find the solution of the equations of (2.8) and (2.4)
in the form of Fourier's integrals
$$
2U_c(x)=\dfrac{1}{2\pi}\int\limits_{-\infty}^{\infty}
e^{ikx}E(k)\,dk,\qquad
\delta(x)=\dfrac{1}{2\pi}\int\limits_{-\infty}^{\infty}
e^{ikx}\,dk,
\eqno{(2.9)}
$$
$$
h_c(x,\mu)=\dfrac{1}{2\pi}\int\limits_{-\infty}^{\infty}
e^{ikx}\Phi(k,\mu)\,dk.
\eqno{(2.10)}
$$

Thus the function $h^+_c(x,\mu)$  is expressed through the spectral density $E(k)$
of mass speed as follows
$$
h_c^+(x,\mu)=-\dfrac{1}{\mu}\exp(-\dfrac{x}{\mu})
\int\limits_{x}^{+\infty} \exp(+\dfrac{t}{\mu})dt
\dfrac{1}{2\pi}
\int\limits_{-\infty}^{+\infty}e^{ikt}E(k,\mu)\,dk=
$$
$$
=\dfrac{1}{2\pi}\int\limits_{-\infty}^{\infty}\dfrac{e^{ikx}
E(k,\mu)}{1+ik\mu}dk.
$$

Similarly we receive that
$$
h_c^-(x,\mu)=\dfrac{1}{2\pi}
\int\limits_{-\infty}^{\infty}\dfrac{e^{ikx}
E(k,\mu)}{1+ik\mu}dk.
$$

Therefore
$$
h_c^{\pm}(x,\mu)=\dfrac{1}{2\pi}
\int\limits_{-\infty}^{\infty}\dfrac{e^{ikx}
E(k,\mu)}{1+ik\mu}dk.
$$

Further we will get with the glance odd parity $E(k)$
$$
h_c^{\pm}(0,\mu)=\dfrac{1}{2\pi}
\int\limits_{-\infty}^{\infty}\dfrac{E(k,\mu)}{1+ik\mu}dk=
\dfrac{1}{2\pi}\int\limits_{-\infty}^{\infty}
\dfrac{E(k)\,dk}{1+k^2\mu^2}=$$$$=
\dfrac{1}{\pi}\int\limits_{0}^{\infty}\dfrac{E(k)\,dk}{1+
k^2\mu^2}.
\eqno{(2.11)}
$$

\begin{center}
{\bf 3. The characteristic equation}
\end{center}

Using the Fourier integral (2.9) and (2.10) the equation (2.4) and (2.11)
is transformed into the following system of equations
$$
E(k)=\dfrac{1}{\sqrt{\pi}}\int\limits_{-\infty}^{\infty}
e^{-t^2}\Phi(k,t)dt,
\eqno{(3.1)}
$$

$$
\Phi(k,\mu)(1+ik\mu)=
$$
$$
=E(k)+|\mu|\Bigg[-2qU_{sl}(q)+G_T\Big(\mu^2-\dfrac{1}{2}\Big)
-\dfrac{q}{\pi}
\int\limits_{0}^{\infty}\dfrac{E(k)dk}{1+k^2\mu^2}\Bigg]
\eqno{(3.2)}
$$

We will express the function $\Phi(k,\mu)$   from equation (3.2)
and we will  substitute it into
the equation (3.1). We get the following characteristic equation
$$
E(k)L(k)=-2qU_{sl}(q)T_1(k)+
$$
$$
+qG_T \Big(T_3(k)-\dfrac{1}{2}T_1(k)\Big)-
\dfrac{q}{\pi}\int\limits_{0}^{\infty}K(k,k_1)E(k_1)dk_1
\eqno{(3.3)}
$$
with the kernel
$$
\dfrac{1}{\sqrt{\pi}}\int\limits_{-\infty}^{\infty}
\dfrac{e^{-t^2}|t|dt}{(1+ikt)(1+k_1^2t^2)}
=$$$$=\dfrac{2}{\sqrt{\pi}}\int\limits_{0}^{\infty}\dfrac{e^{-t^2}t\,dt}
{(1+k^2t^2)(1+k_1^2t^2)}=K(k,k_1).
\eqno{(3.4)}
$$

The equation (3.3) is the Fredholm's  integral equation of the second kind.
In addition, the notation is entered in (3.3)
$$
T_n(k)=\dfrac{2}{\sqrt{\pi}}\int\limits_{0}^{\infty}
\dfrac{e^{-t^2}t^n\,dt}{1+k^2t^2},\quad n=1,2,3,\cdots,
\eqno{(3.5)}
$$
$$
L(k)=1-\dfrac{1}{\sqrt{\pi}}\int\limits_{-\infty}^{\infty}
\dfrac{e^{-t^2}dt}{1+ikt}.
\eqno{(3.6)}
$$
It is easy to see that
$$
L(k)=1-\dfrac{1}{\sqrt{\pi}}
\int\limits_{-\infty}^{\infty}\dfrac{e^{-t^2}dt}{1+k^2t^2}=$$$$=
1-\dfrac{2}{\sqrt{\pi}}\int\limits_{0}^{\infty}\dfrac{e^{-t^2}dt}
{1+k^2t^2}=k^2 \dfrac{2}{\sqrt{\pi}}\int\limits_{0}^{\infty}
\dfrac{e^{-t^2}t^2\;dt}{1+k^2t^2},
$$
or, in short,
$$
L(k)=k^2 T_2(k).
$$

\begin{center}
  \bf 4. The Neumann Series
\end{center}

The solution of equation (3.3) we are looking for in the form
$$
E(k)=qG_T\Big[E_0(k)+q\,E_1(k)+q^2\,E_2(k)+\cdots\Big],
\eqno{(4.1)}
$$
$$
U_{sl}(q)=\dfrac{1}{2}G_T\Big[V_0+V_1q+V_2q^2+\cdots+V_nq^n+\cdots\Big].
\eqno{(4.2)}
$$

Let us substitute the decomposition (4.1) and (4.2) in equation (3.3).
Using the equalities (3.4)-(3.6), we obtain the countable system of equa\-tions
$$
E_0(k)L(k)=-V_0T_1(k)+T_3(k)-\dfrac{1}{2}T_1(k),
\eqno{(4.3)}
$$
$$
E_1(k)L(k)=-V_1T_1(k)-\dfrac{1}{\pi}\int\limits_{0}^{\infty}K(k,k_1)E_0(k_1)dk_1,
\eqno{(4.4)}
$$
$$
E_2(k)L(k)=-V_2T_1(k)-\dfrac{1}{\pi}\int\limits_{0}^{\infty}K(k,k_1)E_1(k_1)dk_1,
\eqno{(4.5)}
$$
$$
............................................................................................
$$
$$
E_n(k)L(k)=$$$$=-V_nT_1(k)-\dfrac{1}{\pi}\int\limits_{0}^{\infty}K(k,k_1)E_{n-1}(k_1)dk_1,
\quad n=1,2,3,\cdots.
\eqno{(4.6)}
$$

From the equation (4.3) we find
$$
E_0(k)=\dfrac{-\Big(V_0+\dfrac{1}{2}\Big)T_1(k)+T_3(k)}{k^2T_2(k)}.
\eqno{(4.7)}
$$

We will eliminate the pole of the second order in right part (4.7).
We will notice that
$$
T_1(k)=\dfrac{1}{\sqrt{\pi}}-k^2T_3(k),\qquad
T_3(k)=\dfrac{1}{\sqrt{\pi}}-k^2T_5(k).
$$

Then
$$
E_0(k)=\dfrac{-\Big(V_0+\dfrac{1}{2}\Big)\dfrac{1}{\sqrt{\pi}}+
\Big(V_0+\dfrac{1}{2}\Big)k^2T_3(k)+\dfrac{1}{\sqrt{\pi}}-k^2T_5(k)}{k^2T_2(k)}.
$$

For elimination of the pole we will demand that
$$
-\Big(V_0+\dfrac{1}{2}\Big)\dfrac{1}{\sqrt{\pi}}+\dfrac{1}{\sqrt{\pi}}=0,
$$
from which we have  $V_0=\dfrac{1}{2}$.  Then
$$
E_0(k)=\dfrac{T_3(k)-T_5(k)}{T_2(k)}.
$$

By means of the last equality from equation (4.4) we find
$$
E_1(k)=-\dfrac{\displaystyle V_1T_1(k)+\dfrac{1}{\pi}\int\limits_{0}^{\infty}
K(k,k_1)E_0(k_1)dk_1}{k^2T_2(k)}.
\eqno{(4.8)}
$$

For the eliminating of the pole in the right part (4.8)   we choose  $V_1$ in the form
$$
V_1=-\dfrac{1}{\sqrt{\pi}}\int\limits_{0}^{\infty}\dfrac{T_1(k_1)}{T_1(0)}E_0(k_1)dk_1=
0.28566.
$$

We will find the numerator of right part (4.8). We have
$$
V_1T_1(k)+\dfrac{1}{\pi}\int\limits_{0}^{\infty}K(k,k_1)E_0(k_1)dk_1=
$$
$$
=\dfrac{1}{\pi}\int\limits_{0}^{\infty}\Big[K(k,k_1)-\dfrac{T_1(k)T_1(k_1)}{T_1(0)}\Big]
E_0(k_1)dk_1.
$$

We notice that
$$
K(k,k_1)=T_1(0)-k_1^2T_3(k_1)-k^2T_3(k)+k^2k_1^2K_3(k,k_1).
$$

We will find the difference
$$
K(k,k_1)-\dfrac{T_1(k)T_1(k_1)}{T_1(0)}=$$$$=K(k,k_1)-
\dfrac{(T_1(0)-k^2T_3(k))(T_1(0)-k_1^2T_3(k_1))}{T_1(0)}=
$$
$$
=k^2k_1^2\Big[K_5(k,k_1)-\sqrt{\pi}T_3(k)T_3(k_1)\Big]=k^2S(k,k_1),
$$
where
$$
S(k,k_1)=k_1^2\Big[K_5(k,k_1)-\sqrt{\pi}T_3(k)T_3(k_1)\Big].
$$

According to (4.8) we get now
$$
E_1(k)=-\dfrac{1}{\pi T_2(k)}\int\limits_{0}^{\infty}S(k,k_1)E_0(k_1)dk_1.
$$

From equation (4.5) we find
$$
E_2(k)=-\dfrac{V_2T_1(k)+\dfrac{1}{\pi}\displaystyle
\int\limits_{0}^{\infty}K(k,k_1)E_1(k_1)dk_1}
{k^2T_2(k)}.
$$

From here we find
$$
V_2=-\dfrac{1}{\sqrt{\pi}}\int\limits_{0}^{\infty}T_1(k_1)E_1(k_1)dk_1=-0.021135.
$$

By means of this correlation we will transform the previous equality  to the form
$$
E_2(k)=-\dfrac{1}{\pi T_2(k)}\int\limits_{0}^{\infty}S(k,k_1)E_1(k_1)dk_1.
$$

Similarly, from equation (4.6), we find
$$
E_n(k)=-\dfrac{v_nT_1(k)+\dfrac{1}{\pi}\displaystyle
\int\limits_{0}^{\infty}K(k,k_1)E_{n-1}(k_1)dk_1}
{k^2T_2(k)}.
$$

It follows that
$$
V_n=-\dfrac{1}{\sqrt{\pi}}\int\limits_{0}^{\infty}T_1(k_1)E_{n-1}(k_1)dk_1, \qquad
n=1,2,3,\cdots.
$$
In this case
$$
E_n(k)=-\dfrac{1}{\pi T_2(k)}\int\limits_{0}^{\infty}S(k,k_1)E_{n-1}(k_1)dk_1,\qquad
n=1,2,3,\cdots.
$$

We will write out formulas for distribution of mass speed in the second  approximation
$$
U_c(x)=U_{sl}(q)+\dfrac{1}{4\pi}\int\limits_{-\infty}^{\infty}e^{ikx}
[E_0(k)+E_1(k)q+E_2(k)q^2+\cdots]dk.
$$

For the function $h(x,\mu)$  we get
$$
h(x,\mu)=2U_{sl}(q)-G_T\Big(\mu^2-\dfrac{1}{2}\Big)+\dfrac{1}{2\pi}
\int\limits_{-\infty}^{\infty}e^{ikx}\Phi(k,\mu)dk,
$$
where
$$
\Phi(k,\mu)=\dfrac{1-ik\mu}{1+k^2\mu^2}\Bigg[E_0(k)+q\Bigg(E_1(k)+|\mu|G_T
\Big(\mu^2-\dfrac{1}{2}-V_0\Big)-
$$
$$
-|\mu|\dfrac{1}{\pi}\int\limits_{0}^{\infty}\dfrac{E_0(k_1)dk_1}{1+k_1^2\mu^2}\Bigg)+
q^2\Bigg(E_2(k)-|\mu|G_TV_1-\dfrac{|\mu|}{\pi}\int\limits_{0}^{\infty}
\dfrac{E_1(k_1)dk_1}{1+k_1^2\mu^2}\Bigg)\Bigg].
$$

\begin{center}
  \bf 5. Discussion of results and conclusions
\end{center}

We will compare the got results to previous. In particular, with the precise
decision for the speed of thermal sliding at $q=1$: $U_{sl}=0.38316 G_T$
(see \cite{12}).

We will rewrite this formula in the  dimensional  form: $u_{sl}=1.1495\zeta g_T$.

Here $\zeta$ is the kinematic  viscosity of gas.

According to the got results  the dimensionless speed of the thermal  sliding is equal
$$
U_{sl}(q)=\dfrac{1}{2}(V_0+V_1q+V_2q^2)G_T=
0.5(0.5+0.2857q-0.0211q^2)G_T.
$$

We will bring this formula over to the  dimensional  form
$$
u_{sl}(q)=0.75(1+0.5714q-0.0422q^2)\zeta g_T.
\eqno{(5.1)}
$$

In the zero approximation from a formula (5.1) it is visible that we received
exact result of Maxwell for  mirror boundary conditions: $u_{sl}=0.75\zeta g_T$.

We will  introduce the relative error
$$
O_n(q)=\dfrac{u_{sl}-u_{sl}(q)}{u_{sl}}\cdot 100\%,
\eqno{(5.2)}
$$
where $u_{sl}(q)$ is determined by the equality (5.1).

In the first approximation   the speed of the thermal sliding is equal
$$
u^{(1)}_{sl}(q)=0.75(0.5+0.2857q)\zeta g_T,
$$
in the second approximation it is determined by equality (5.1)
$$
u_{sl}^{(2)}(q)=0.75(1+0.5714q-0.0422q^2)\zeta g_T.
$$

It is easy to see that the relative error is equal in a zero approximation
$O_0(1)=34.48\%$, in the first it is equal $O_1(1)=2.52\%$,
in the second it is equal $O_2(1)=0.23\%$.

Comparing the brought estimations over, we come to the conclusion that exactly
the nonlinear analysis leads to an efficient approximating formula for the calculation
of  the speed of  the thermal sliding.

We will rewrite a formula (5.1) to the form $u_{sl}(q)=K(q)\zeta g_T$, where $K(q)$
is the coeffi\-ci\-ent of the thermal  sliding.
Let us compare our coef\-fi\-ci\-ent  from  this work with the coefficients found in
\cite{6}-\cite{8}.
The same coefficient was found in works \cite{6} and \cite{7}: $K(q)=0.75(1+0.5q)$.
In work \cite{8} it was obtained that $K(q)=0.75(0.5321+q)$.
The comparison of this coefficient with (5.1) according to (5.2) shows that at $q=1$
the coefficient deviation from \cite{6} and \cite{7} does not exceed $0.2\%$ from the
coefficient received in work.

Thus, in work efficient approximating formulas for the solution of the
problem on thermal sliding with mirror-diffuse boundary conditions are removed.
Further authors intend to consider the problem of thermal sliding for quantum gases.

\end{document}